\documentclass[prb,showpacs,preprint,amsmath,amssymb,superscriptaddress]{revtex4}
\usepackage{epsfig}
\usepackage{graphicx}
\usepackage{dcolumn}
\usepackage{bm}

\begin{document}
\title{{\it Ab~initio} exchange interactions and magnetic properties of Gd$_2$Fe$_{17}$ iron sublattice:
rhombohedral vs. hexagonal phases}
\author{A.V.~Lukoyanov}
\affiliation{Institute of Metal Physics, Russian Academy of
Sciences-Ural Division, 620041 Yekaterinburg, Russia}
\affiliation{Ural State Technical University-UPI,
620002 Yekaterinburg, Russia}
\author{E.E.~Kokorina}
\affiliation{Institute of Electrophysics, Russian Academy of
Sciences-Ural Division, 620016 Yekaterinburg, Russia}
\author{M.V.~Medvedev}
\affiliation{Institute of Electrophysics, Russian Academy of
Sciences-Ural Division, 620016 Yekaterinburg, Russia}
\author{I.A.~Nekrasov}
\affiliation{Institute of Electrophysics, Russian Academy of
Sciences-Ural Division, 620016 Yekaterinburg, Russia}

\date{\today}

\begin{abstract} 

In the framework of the LSDA+$U$ method electronic structure
and magnetic properties of the intermetallic 
compound Gd$_2$Fe$_{17}$ for both rhombohedral and hexagonal phases have been calculated.
On top of that, {\it ab~initio} exchange interaction parameters within the Fe sublattice
for all present nearest and next nearest Fe ions have been obtained.
It was found that for the first coordination sphere exchange interaction
is ferromagnetic. For the second coordination sphere exchange interaction
is observed to be weaker and of antiferromagnetic type. 
Employing the theoretical values of exchange parameters Curie temperatures $T_C$
for the first as well as for the first and second coordination spheres
of both hexagonal and rhombohedral phases of Gd$_2$Fe$_{17}$
within Weiss mean-field theory were estimated.
Obtained values of $T_C$ and its increase going from the hexagonal to rhombohedral
crystal structure of Gd$_2$Fe$_{17}$ agree well with experiment.
Also for both structures LSDA+$U$ computed values of total magnetic moment
agree well with experimental ones.

\end{abstract}


\maketitle 
\section{Introduction}

Intermetallic compounds $R_2$Fe$_{17}$ with high concentration of iron 
($R$ is a rare-earth ion) are of interest for both experimental and 
theoretical investigations.~\cite{Kou98} These compounds are distinct 
for a relatively large magnetic moment -- mainly caused by the magnetic moments 
at the Fe ions -- and comparatively low Curie temperature $T_C$. 
The compounds are crystallized in two related crystal structures -- 
rhombohedral $Th_{2}Zn_{17}$-type and hexagonal $Th_{2}Ni_{17}$-type.
In each of these structures, the Fe ions can be 
located in four inequivalent crystallographic positions, and for this 
reason various experiments reveal four values of the local magnetic 
moments at different positions of the Fe ions. Also, most of the 
intermetallic compounds $R_2$Fe$_{17}$ have collinear magnetic structure 
except for Ce$_2$Fe$_{17}$, Tm$_2$Fe$_{17}$, and Lu$_2$Fe$_{17}$,~\cite{Buschow97} 
where the magnetic moments of the Fe ions in all crystallographic positions 
are collinear to each other, but frequently antiparallel 
to the magnetic moments of the rare-earth ions $R$.

Initially, the moderate Curie temperatures $T_C$ in the compounds 
$R_2$Fe$_{17}$ was explained on the base of assumption\cite{Givord74} 
of a critical distance $r_c\approx2.5$~\AA$~$between two ions 
in a Fe-Fe pair, so that for the distances $r < r_c$ the exchange 
interaction is antiferromagnetic, whereas for $r > r_c$ -- ferromagnetic. 
From this assumption it followed that a low value of $T_C$ is caused 
by a competition of ferromagnetic and antiferromagnetic exchange 
interactions between neighboring pairs of the Fe ions in various 
crystallographic positions. It also meant that an expansion 
of the lattice should result in an increased number of ferromagnetic 
interactions and corresponding growth of $T_C$. Such an expansion 
of the lattice can be reached either inserting nonmagnetic interstitial 
atoms,~\cite{Sun90,Mooij88} or substituting the Fe ions by nonmagnetic 
atoms with the larger ionic radii.~\cite{Jacobs92,Shen93}

However, in a number of experiments, for example, for substitutional 
solid solutions $R_2$Fe$_{17-x}$Si$_x$, it was found that the crystal 
lattice contracts, and $T_C$ grows.~\cite{vanMens86,Li95,Kuchin98} 
These results question the adequacy of the simple model of exchange 
interactions of neighboring exchange-bounded ions that change the sign 
at the critical distance $r_c$ between the ions.

To determine a relation between $T_C$ and exchange 
interactions in the $R_2$Fe$_{17}$ series, the compound Gd$_2$Fe$_{17}$ 
is of especial interest. First, among the series Gd$_2$Fe$_{17}$ 
has the highest Curie temperature $T_C$. Secondly, Gd itself is located 
at the middle of the rare-earth series, and Gd$_2$Fe$_{17}$ separates 
the compounds with light rare-earth elements crystallizing 
in the rhombohedral $Th_2Zn_{17}$-type structure, from the compounds 
with heavy rare-earth elements crystallizing in the hexagonal 
$Th_2Ni_{17}$-type structure.
Thus Gd$_2$Fe$_{17}$ is an intriguing example among
the $R_2$Fe$_{17}$ series which can be found in both crystal 
structures.

Apparently, in most experiments samples are mixture of two structures. 
At the same time, there are experimental works for either pure rhombohedral 
structure,~\cite{Shen98} or pure hexagonal structure\cite{Knyazev06} 
of Gd$_2$Fe$_{17}$. For Gd$_2$Fe$_{17}$ in the rhombohedral structure, 
the values of critical temperature and saturation magnetization 
were found to be $T_C^{rh}$=475~K,~\cite{Shen98} 
$M_S^{rh}$=21.5~$\mu_B$/f.u.,~\cite{Shen98} respectively. 
For the hexagonal structure, these values are $T_C^{hex}$=466~K 
and $M_S^{hex}$=21.2~$\mu_B$/f.u.~\cite{Knyazev06} 
Although these two sets of experimental data differ slightly and further 
experimental study is needed, one can notice that Gd$_2$Fe$_{17}$ 
in the rhombohedral phase has the higher values of magnetic characteristics, 
than in the hexagonal one.

Motivated by this fact, in the paper we consider 
how the transfer from the rhombohedral to hexagonal structures 
in Gd$_2$Fe$_{17}$ influences the values of the local magnetic moments 
and exchange interaction parameters. The later values are obtained using 
the results of electronic structure calculations within LSDA+$U$ method.
Further values of Curie temperature  were estimated in the framework of the
Weiss mean-field theory. Moreover based on these calculations
we propose explanation of rather low $T_C$ values and relatively
small saturation magnetization of $R_2$Fe$_{17}$ series
with respect to elemental $bcc$ Fe.

\section{
Crystal structures and local magnetic moments}
\label{moments}

The intermetallic compound Gd$_2$Fe$_{17}$ can crystallize both 
in the rhombohedral $Th_2Zn_{17}$-type structure (space group R$\bar{3}$m 
-- no. 166 in International Tables for Crystallography) and the hexagonal 
$Th_2Ni_{17}$-type crystal structure (space group P6$_3$/mmc -- no. 194). 
Main structural blocks are iron hexagons building layers. Most of these 
hexagons have empty centers, whereas others contain the rare-earth 
Gd ions or mediate the Fe ions in dumbbell positions.

\begin{figure}[t]
\begin{center}
\epsfxsize=14cm
\epsfbox{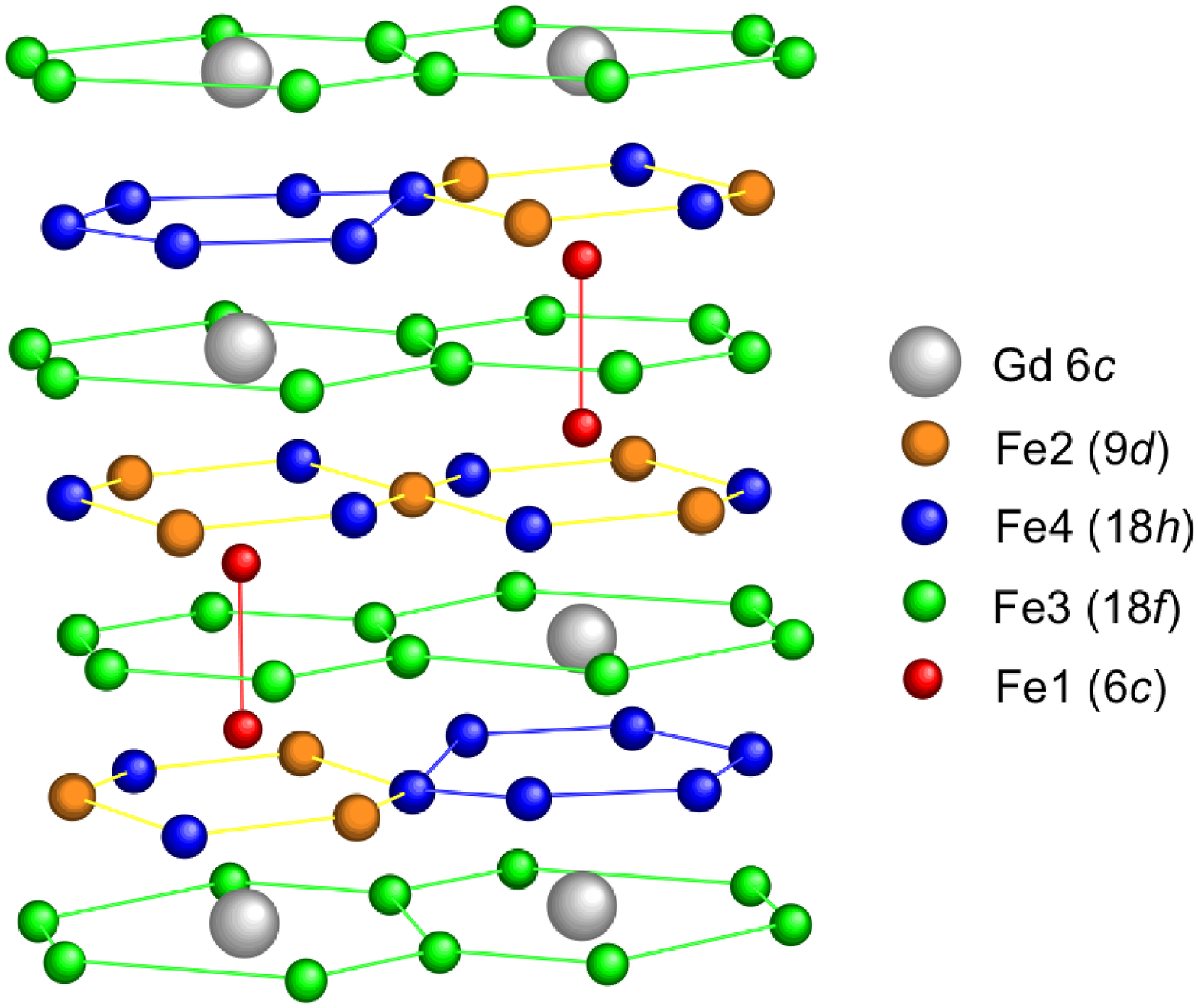}
\end{center}
\caption{(Color online) 
Rhombohedral $Th_2Zn_{17}$-type structure of Gd$_2$Fe$_{17}$.}
\label{fig1}
\end{figure}

The rhombohedral structure of the $Th_2Zn_{17}$-type is shown 
in Fig.~\ref{fig1}. In this type of structure the Fe(18$f$) hexagons 
(let us denote Fe(18$f$) as Fe3 for the rhombohedral structure) 
contain the Gd ions in positions 6$c$ either at the center of each hexagon, 
as it is presented in the upper and lower layers in Fig.~\ref{fig1}, 
or alternate with empty hexagons, as it is shown in the inner two layers. 
Interlayer Fe(6$c$)=Fe1 ions at the dumbbell positions lower the symmetry 
of some iron ions in the nearest intermediate layers from 18$h$ to 9$d$ 
(above and below the dumbbell), and hence, these intermediate layers 
without Gd contain ions Fe(9$d$)=Fe2, as well as Fe(18$h$)=Fe4 
in the hexagons. Unit cell of the rhombohedral structure consists 
of one formula unit of Gd$_2$Fe$_{17}$.

\begin{figure}[t]
\begin{center}
\epsfxsize=14cm
\epsfbox{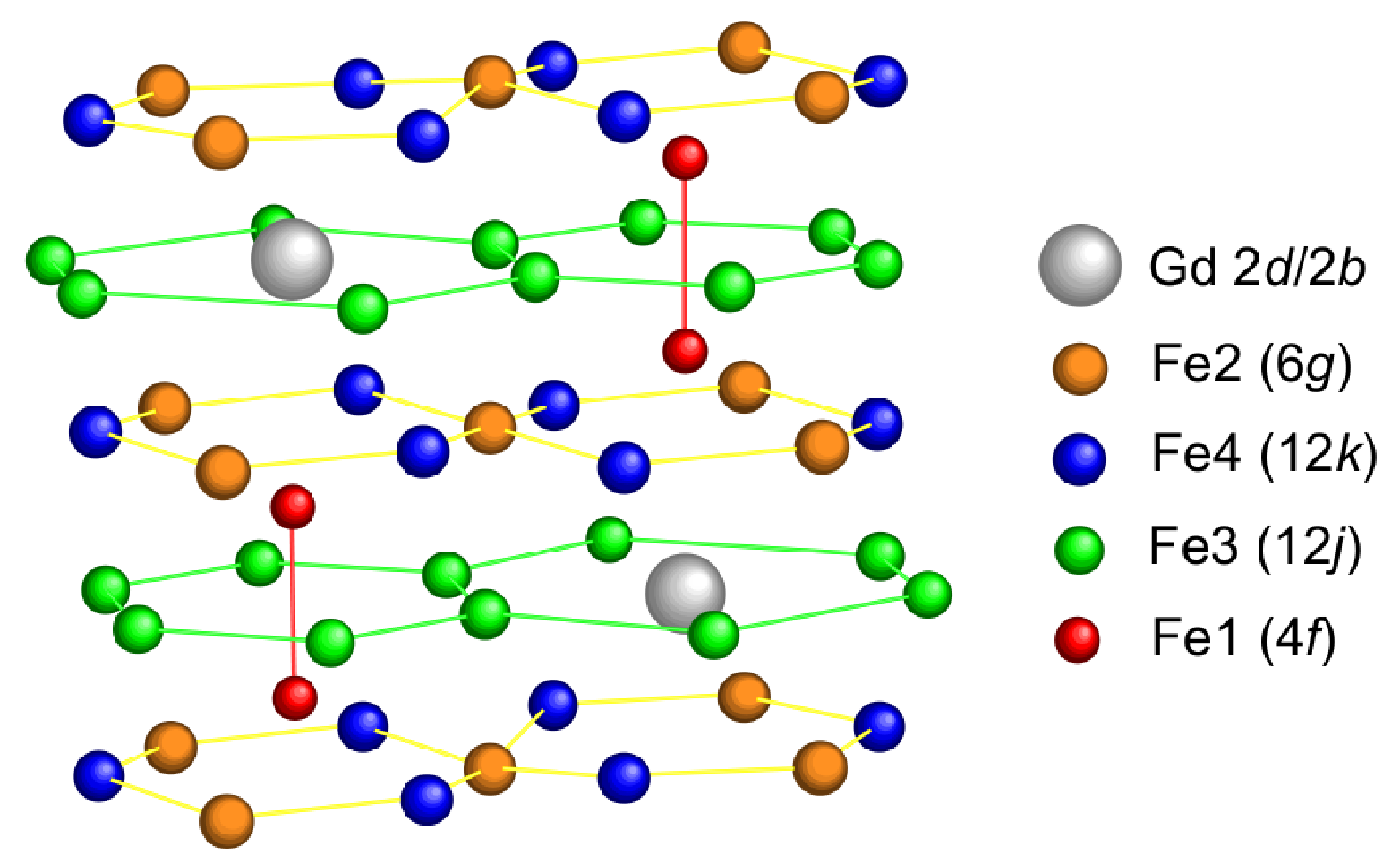}
\end{center}
\caption{(Color online) 
Hexagonal $Th_2Ni_{17}$-type structure of Gd$_2$Fe$_{17}$.}
\label{fig2}
\end{figure}

The hexagonal structure of the $Th_2Ni_{17}$-type is shown in Fig.~\ref{fig2}. 
In contrast with the rhombohedral structure, there are no layers 
where each hexagon contains the Gd ion. In this structure the Gd ions 
are located only at the centers of the hexagons (let us denote 
Fe(12$j$)=Fe3 for the hexagonal structure) alternating with 
the hexagons with the dumbbells of Fe(4$f$) (Fe(4$f$)=Fe1), 
and the layers with the hexagons containing Gd at each center are absent. 
Moreover, here one can find the alternating intermediate layers Fe(6$g$)=Fe2 
and Fe(12$k$)=Fe4 without the Gd ions. Unit cell of this structure 
consists of two formula units of Gd$_2$Fe$_{17}$ due to the different 
symmetry of the rare-earth ions coordination environment 
in the $ab$-plane translations.

For both structures in the layers with alternating Gd-containing and 
empty hexagons Fe3, these hexagons are deformed -- the hexagons with 
Gd are slightly expanded, the hexagons without Gd -- contracted. 
Furthermore, because for the hexagons Fe2 and Fe4 the intermediate 
layers the Gd ion and the dumbbell Fe1 ions are alternating from 
one layer to another, so the ions Fe2 and Fe4 in these layers 
are slightly distorted upward and downward with respect 
to the plane, i.e. these layers are slightly corrugated.

The electronic structure of the rhombohedral structure 
of Gd$_2$Fe$_{17}$ was calculated in the LSDA+$U$ method 
(Ref.~\onlinecite{Anisimov97}) in the framework of the band 
calculation package TB-LMTO-ASA (tight-binding, linear 
muffin-tin orbitals, atomic sphere approximation).~\cite{Andersen75} 
However let us mention here that only Gd 4f-shell was treated within the
LSDA+$U$ method since correlation effects are much stronger there than for
the Fe 3d-shell ($U_{Gd}\sim$7~eV (see below), $U_{Fe}\sim$2~eV\cite{Anisimov91,Lichtenstein01}).
We believe that greater part (not all) of 
electronic interaction effects is taken into consideration
for Fe 3d-shell in the frame of LSDA approximation. 
It is well enough justified since responsible Stoner parameter for Fe 3d-shell
$J^S_{Fe}\sim$1~eV is close to $U_{Fe}$.
For the hexagonal structure, the results of such a calculation 
were reported in Ref.~\onlinecite{Knyazev06}, below we include 
some details of that calculation to provide the full picture 
for comparison of the rhombohedral and hexagonal structures 
of Gd$_2$Fe$_{17}$. Lattice parameters for the rhombohedral structure 
are $a$=8.538~$\AA$ and $c$=12.431~$\AA$ (Ref.~\onlinecite{Buschow77}), 
for the hexagonal structure -- $a$=8.496~$\AA$ and $c$=8.341~$\AA$ 
(Ref.~\onlinecite{Knyazev06}). Atomic spheres radii were chosen 
as $R$(Gd)=3.72~a.u. and $R$(Fe)=2.66~a.u. for the hexagonal structure, 
and $R$(Gd)=2.86~a.u.\cite{Gd_radii} and $R$(Fe)=2.62~a.u. -- for the rhombohedral 
structure.~\cite{Knyazev06} Orbital basis contained 6$s$, 6$p$, 5$d$, 
and 4$f$ muffin-tin orbitals for Gd and 4$s$, 4$p$, and 3$d$ 
for Fe sites. Integration over the first Brillouin zone was performed 
using 32 irreducible {\bf k}-points (6$\times$6$\times$6=216 is a total 
number of {\bf k}-points). Empty atomic spheres without nuclear charge 
were inserted in the case of the hexagonal structure to fill 
the interstitial regions.

In the LSDA+$U$ method Coulomb interaction for Gd 4f-orbitals was taken into account 
via parameters of direct $U$ and exchange $J$ Coulomb interactions 
of the $4f$-electrons in Gd. For Gd$_2$Fe$_{17}$ in both structures 
the calculations of these parameters in the constrained 
LDA method\cite{Gunnarsson89} resulted in $U_{Gd}$=6.7~eV 
and $J_{Gd}$=0.7~eV that is in agreement with the values 
for Gd metal.~\cite{Anisimov97}

Local magnetic moments for various Fe and Gd sites obtained 
in the calculations are listed in Table~\ref{tab1} 
for both phases under consideration. Note that in qualitative 
discussions of magnetic properties of $R_2$Fe$_{17}$ especial 
attention is frequently paid to the iron ions in the dumbbell 
positions (Fe1 in our notations). From Table~\ref{tab1} 
one can see that magnetic moments 
of the Fe1 ions for the hexagonal structure are larger than 
for the rhombohedral one.
However, total magnetic moment $M$($\Sigma$Fe) value
of the iron subsystem in the rhombohedral phase is larger 
($M_{rh}$($\Sigma$Fe)=38.04~$\mu_B$/f.u. vs. 
$M_{hex}$($\Sigma$Fe)=36.54~$\mu_B$/f.u..
Experimental value of total Fe ions magnetic moment
for the mixture of two phases is 36.9~$\mu_B$/f.u.).\cite{Liu94}
Magnetic moments of the Gd ions in both calculations are antiparallel 
to the iron magnetic moments, so that the total magnetic 
moments are $M_{rh}$($\Sigma$Fe + Gd)=23.70~$\mu_B$/f.u. 
and $M_{hex}$($\Sigma$Fe + Gd)=22.21~$\mu_B$/f.u., respectively.
These values agree well with experimental ones
$M_S^{rh}$=21.5~$\mu_B$/f.u.~\cite{Shen98} 
and $M_S^{hex}$=21.2~$\mu_B$/f.u.,~\cite{Knyazev06} correspondingly. 

\begin{table*}
\caption{Calculated values of local magnetic moments 
for both rhombohedral and hexagonal phases of Gd$_2$Fe$_{17}$.}
\begin{ruledtabular}
\begin{tabular}{ldld}
\multicolumn{2}{c}{rhombohedral structure}& \multicolumn{2}{c}{hexagonal structure}\\
Site & \mbox{$M(\mu_B)$} & Site & \mbox{$M(\mu_B)$} \\
\hline
Gd($6c$)     & -7.17 & Gd($2b$) &  -7.13 \\
~~~~~~~~     & ~~~~~ & Gd($2d$) &  -7.20 \\
Fe($6c$)=Fe1 & 2.19 & Fe($4f$)=Fe1 &  2.31 \\
Fe($9d$)=Fe2 & 2.26 & Fe($6g$)=Fe2 &  2.10 \\
Fe($18f$)=Fe3 & 2.17 & Fe($12j$)=Fe3 &  2.40 \\
Fe($18h$)=Fe4 & 2.31 & Fe($12k$)=Fe4 &  1.87

\end{tabular}
\end{ruledtabular}
\label{tab1}
\end{table*}

\section{
Exchange interactions}
\label{exch_int}

In the framework of spin-fluctuation theory of magnetism 
it was found\cite{Hubbard79,Prange79,Wang82,Mathon83} 
that the magnetic behavior of transition metals can be 
described by the Heisenberg model with long-range exchange 
interactions between classical spin vectors, in general, 
their magnitudes are non-integer. For an intermetallic magnet 
with several types of magnetic moments a spin Hamiltonian 
of such a model can be written as 

\begin{equation}
H=-\frac{1}{2} \sum_{a, b}\sum_{\mathbf{l}_a, 
\mathbf{n}_b, \mathbf{l}_a \neq \mathbf{n}_b} 
I_{ab}(\mathbf{l}_a - \mathbf{n}_b) \mathbf{S}_a(\mathbf{l}_a) 
\mathbf{S}_b(\mathbf{n}_b),
\label{Hamiltonian}
\end{equation}
where $a$ and $b$ are types of magnetic ions 
(e.g. $a$=Fe1, Fe2, Fe3, and Fe4), $\mathbf{l}_a$ 
-- radius vector of a magnetic ion of the $a$ type, 
$\mathbf{S}_a(\mathbf{l}_a)$ -- classical spin vector 
with the magnitude $S_a$, corresponding to magnetic 
moment $\mu_a=g \mu_B S_a$, and 
$I_{ab}(\mathbf{l}_a - \mathbf{n}_b)$ is an exchange 
parameter between ions of the types $a$ and $b$ 
with the distance $\mathbf{l}_a - \mathbf{n}_b$ between 
the ions.

Later an ab-initio method to calculate the exchange interaction 
parameters between different lattice sites for the classical
Heisenberg model at $T$=0 was proposed.~\cite{Liechtenstein84} 
In that scheme the exchange interaction parameters are determined 
calculating the second-order derivative of the total energy 
at $T$=0 with respect to small deviations of magnetic moments 
of the corresponding lattice sites from collinear magnetic configuration. 
This method was used in the present work to calculate 
the exchange interaction parameters $I_{ab}$ ($a$, $b$=Fe1, Fe2, 
Fe3, Fe4) in the iron sublattice of both structures of Gd$_2$Fe$_{17}$.

First note that in Ref.~\onlinecite{Liechtenstein84} the energy 
of exchange interaction between different sites of the magnetic 
lattice is presented in the form
$E_{ex}(\mathbf{l}-\mathbf{n})=-2J^L_{ab}
(\mathbf{l}_a-\mathbf{n}_b)
\mathbf{e}_a(\mathbf{l}_a)
\mathbf{e}_b(\mathbf{n}_b)$
as an exchange between two classical unit spin 
vectors $\mathbf{e}_a(\mathbf{l}_a)=(sin \phi_a(\mathbf{l}_a) sin 
\Theta_a(\mathbf{l}_a), cos \phi_a(\mathbf{l}_a) 
sin \Theta_a(\mathbf{l}_a), cos \Theta_a(\mathbf{l}_a))$ 
and $\mathbf{e}_b(\mathbf{n}_b)$ regardless of the magnitude 
of the magnetic moments $\mu_a$ and $\mu_b$ at the sites 
$\mathbf{l}_a$ and $\mathbf{n}_b$ under consideration. 
Since the classical spin vector $\mathbf{S}_a(\mathbf{l}_a)$ 
is equal to $\mathbf{S}_a(\mathbf{l}_a)=S_a \mathbf{e}_a(l_a)$, 
hence the relation between the exchange parameters $J_{ab}^{L}$ 
calculated using the method of Ref.~\onlinecite{Liechtenstein84} 
and the exchange parameters $I_{ab}$ of the Hamiltonian (\ref{Hamiltonian}) 
will be given as $2J_{ab}^{L}=I_{ab} S_a S_b$. Moreover, 
some authors\cite{Mazurenko06} calculating exchange parameters
for insulators prefer to write the energy of exchange 
interaction as 
$E_{ex}(\mathbf{l}-\mathbf{n})=-J^M_{ab}
(\mathbf{l}_a-\mathbf{n}_b)
\mathbf{S}_a(\mathbf{l}_a)
\mathbf{S}_b(\mathbf{n}_b)$
via the exchange parameters $J_{ab}^M$ between classical
spin vectors with the magnitude $S_a,S_b$=1/2. It is easy to demonstrate 
that if one uses the method of Ref.~\onlinecite{Mazurenko06} 
to calculate $J_{ab}^M$, then our exchange parameters $I_{ab}$ 
are related with $J_{ab}^M$ via the equality $I_{ab}=J_{ab}^M/(4 S_a S_b)$.

Generally, in the intermetallic compounds of the $R_2$Fe$_{17}$ 
type there are many types of exchange interaction: $R$-$R$, $R$-Fe, 
and Fe-Fe. The weakest interaction is an indirect interaction 
of the $R$-$R$ type between the rare-earth ions, because even for 
the nearest rare-earth ions their wave functions practically do not overlap.

Exchange interaction $R$-Fe is also indirect since direct overlap 
of $4f$ orbitals of the $R$ ions and $3d$ orbitals of the neighboring 
iron ion is absent, and only small overlap of a polarized $5d$ 
orbital of the rare-earth ion and $3d$ orbital of the iron ions 
is possible. Theoretical estimation of this interaction is still 
difficult, nevertheless, from some indirect experiments this 
interaction can be estimated as 10--30 meV.~\cite{Liu94} 
The sign of the exchange interaction $R$-Fe defines 
the orientation of the magnetic moment on the rare-earth 
ion with respect to the magnetic moments of the iron sublattice 
that seriously affects the saturation moment. At the same time, 
the dominating contribution to free energy of the magnetic system 
is due to the interaction of the Fe-Fe type which 
defines in fact the value of the critical temperature $T_C$. 
Because the number of iron ions is 8.5 times greater
than the number of $R$ ions, and 
in the Fe subsystem -- there is also direct exchange between 
the neighboring Fe ions and the overlapping $3d$ wave functions.

In such complicated crystall structures as $Th_2Zn_{17}$ 
and $Th_2Ni_{17}$ the nearest neighbors in different crystallographic 
directions are arranged at different distances. However, 
analyzing the nearest neighbors of magnetic ions of each type 
Fe1, Fe2, Fe3, and Fe4 in both structures, the following 
regularities can be found. If one considers a spherical layer 
with the radii from $r^{rh}_1$=2.385~$\AA < r < r^{rh}_2$=2.740~$\AA$ 
around a particular ion Fe$a$ ($a$=1, 2, 3, 4) 
in the rhombohedral structure, then all nearest magnetic 
neighbors will be within this coordination layer. Next is 
an empty spherical layer without ions 
$r^{rh}_2$=2.740~$\AA < r < r^{rh}_3$=4.044~$\AA$. 
The second spherical coordination layer will be at 
4.044~\AA~$< r^{rh}_3 < $ 4.3~$\AA$.~\cite{2nd_layer}
The only exception from 
this regularity is the Fe3 class with the nearest neighbor 
Fe3-type ion at $r^{rh}_{in}$=3.563~$\AA$ between the first 
and second coordination layers.
It is typical that the atomic sphere of the chosen (``central'') ion 
Fe$a$ ($a$=1, 2, 3, 4) and atomic spheres 
of nearest Fe$a$ ($a$=1, 2, 3, 4) ions overlap.
Therefore the direct exchange gives a dominant contribution
in exchange parameters of the nearest neighbor Fe ions.

Similarly, in the hexagonal structure of Gd$_2$Fe$_{17}$ 
each ion Fe$a$ ($a$=1, 2, 3, 4) can be surrounded by the first 
spherical coordination layer with the radii 
$r^{hex}_1$=2.399~$\AA < r < r^{hex}_2$=2.735~$\AA$, 
then an empty spherical layer without ions 
with $r^{hex}_2$=2.735~$\AA < r < r^{hex}_3$=4.108~$\AA$, 
and the second spherical coordination layer with
4.108 $\AA$ < $r^{hex}_3 $ <  4.3~$\AA$~\cite{2nd_layer}.
In exactly the same way the Fe3 ions 
violate this regulation, since these ions have a neighbor 
at $r^{hex}_{in}$=3.605~$\AA$ between the first and second 
coordination layers.

The results for the exchange interaction parameters 
$I_{ab}(\mathbf{l}_a - \mathbf{n}_b)$ between different 
pairs of iron ions (in units of K) are presented in 
Tables~\ref{tab2}--\ref{tab5}. 
In Table~\ref{tab2} the parameters of exchange interaction 
(in the descending order of absolute values) between 
the central ion Fe$a$ ($a$=1, 2, 3, 4)  and ions of different 
types from the first coordination layer are presented. 
Here $I_{34}(1)$ stands for the value of the exchange parameter 
between ions Fe3 and Fe4 at the nearest distance between 
these ions (index 1 in the brackets), then $I_{34}(2)$ denotes 
the value of the exchange parameter between ions Fe3 and Fe4 
at the second-order distance between these ions (index 2). 
Moreover, in Table~\ref{tab2} the number of neighbors $Z_{ab}(r)$ 
for a given distance are presented. This value gives 
the number of neighbors for the central ion Fe$a$ 
at the distance $r$. One should note that in our crystal 
structures $Z_{ab}(r) \neq Z_{ba}(r)$ (e.g. $Z_{13}(1)$=6 
and  $Z_{31}(r)$=2, i.e. the Fe1 ion has 6 neighbors of Fe3 
at $r_{13}(1)$=2.735~$\AA$, whereas the Fe3 ion has only 2 Fe1 
ions at the same distance $r_{31}(1)$=2.735~$\AA$).

\begin{table*}
\caption{Parameters of exchange in the hexagonal 
structure of Gd$_2$Fe$_{17}$ for the ions of the first coordination sphere.}
\begin{ruledtabular}
\begin{tabular}{ccccc}
N & Exchange (K) & Distance ($\AA$) & Number of neighbors & Type\\
\hline
1 & $I_{11}$(1)=238.8 & $r_{11}$(1)=2.400 & $z_{11}$(1)=1 & 
Fe1 (dumbbell) - Fe1 (dumbbell)\\
2 & $I_{44}$(1)=218.5 & $r_{44}$(1)=2.477 & $z_{44}$(1)=2 & 
Fe4 (corrugated plane) - Fe4 (corrugated plane)\\
3 & $I_{34}$(1)=136.1 & $r_{34}$(1)=2.511 & $z_{34}$(1)=$z_{43}$(1)=2 & 
Fe3 (dumbbell) - Fe4 (dumbbell)\\
4 & $I_{33}$(2)=123.4 & $r_{33}$(2)=2.512 & $z_{33}$(2)=1 & 
Fe3 - Fe3 \\
5 & $I_{24}$(1)=101.9 & $r_{24}$(1)=2.459 & $z_{24}$(1)=4, $z_{42}$(1)=2 & 
Fe2 - Fe4 (corrugated layer)\\
6 & $I_{23}$(1)=90.7 & $r_{23}$(1)=2.464 & $z_{23}$(1)=4, $z_{32}$(1)=2 & 
Fe2 (corrugated layer) - Fe3 (upper and lower layer)\\
7 & $I_{13}$(1)=90.5 & $r_{13}$(1)=2.735 & $z_{13}$(1)=6, $z_{31}$(1)=2 & 
Fe1 (dumbbell) or Fe3 - Fe1 (dumbbell above and below)\\
8 & $I_{33}$(1)=80.4 & $r_{33}$(1)=2.403 & $z_{33}$(1)=1 & 
Fe3 - Fe3\\
9 & $I_{14}$(1)=72.0 & $r_{14}$(1)=2.653 & $z_{14}$(1)=3, $z_{41}$(1)=1 & 
Fe1 (dumbbell) - Fe4 (nearest corrugated layer)\\
10& $I_{12}$(1)=69.3 & $r_{12}$(1)=2.616 & $z_{12}$(1)=3, $z_{21}$(1)=2 & 
Fe1 (dumbbell) - Fe2 (nearest corrugated layer)\\
11& $I_{34}$(2)=61.3 & $r_{34}$(2)=2.672 & $z_{34}$(2)=$z_{43}$(2)=2 & 
Fe3 - Fe4 (corrugated layer above and below)\\
\end{tabular}
\end{ruledtabular}
\label{tab2}
\end{table*}

It is observed that for the hexagonal Gd$_2$Fe$_{17}$ all exchange 
parameters for each Fe$a$ ion with other ions Fe from 
the first spherical layer are positive i.e. ferromagnetic. 
Surprisingly, the exchange interaction parameter between 
the Fe4 ions within the corrugated hexagon plane 
$I_{44}$(1)=218.5~K is just a little less than 
the exchange parameter $I_{11}$(1)=238.8~K between 
the Fe1 ions at the dumbbell positions. Also, one should 
note that there is no simple dependence between the distance of
the interacting Fe ions and the value of exchange interaction. 
For example, in the pair Fe3-Fe3 at the second-order 
distance $r_{33}$(2)=2.512~$\AA$ the exchange interaction 
parameter is equal to $I_{33}$(2)=123.4~K, whereas 
in the same plane of hexagons Fe3 in the pair Fe3-Fe3 
at the distance $r_{33}(2)$=2.403~$\AA$ the exchange 
interaction parameter is smaller, namely $I_{33}$(1)=80.4~K.

The method of exchange interaction parameters calculation employed here
works in such a way that does not distinguish different mechanisms.
There are several exchange interaction scenario
super exchange, double exchange, direct exchange and RKKY one.
First two require nonmagnetic interstitial ions in the system,
that is not the case here.
Thus only last two mechanisms are applicable.
Direct exchange mostly come from orbitals overlap.
One can suppose that absence of straightforward correlations
between exchange parameter values and corresponding interatomic distances
for the first coordination sphere for different crystallographic directions is
connected with the way how Fe-3d orbitals overlap in these directions.
However classification of the orbitals for these quite complicated
crystal structures is highly non-trivial task which we postpone for future
investigations.
In addition some anisotropy in the first coordination sphere of exchange parameters might come
from RKKY interaction since the Fermi surface of these compounds
is rather complicated.
For the second coordination sphere RKKY interaction is dominant but
still might be highly anisotropic because of direction dependence of ${\bf k}_F$.

In Table~\ref{tab3} the parameters of exchange interaction 
of the ion Fe$a$ ($a$=1, 2, 3, 4) with some ions from 
the second-order coordination layer with 
$r > r_3^{hex}$=4.108~$\AA$, as well as the exchange 
parameters $I_{33}(3)$ for $r_{33}(3)=r_{in}^{hex}$=3.605~$\AA$ 
in a Fe3-Fe3 pair at the third-order distance, 
in the hexagonal structure of Gd$_2$Fe$_{17}$ are presented. 
At these distances 
the exchange parameters are mostly negative, antiferromagnetic, 
and are usually an order of magnitude less than the exchange parameters 
with the ions of the first-order coordination layer. 
At the same time, a sufficiently strong antiferromagnetic 
exchange $I_{22}(1)$=--111.1~K in both pairs 
of the nearest neighbors Fe2-Fe2 from the neighboring 
corrugated planes is present.

\begin{table*}
\caption{Parameters of indirect exchange in the hexagonal 
structure of Gd$_2$Fe$_{17}$ for ions of the second coordination sphere.}
\begin{ruledtabular}
\begin{tabular}{ccccc}
N & Exchange (K) & Distance ($\AA$) & Number of neighbors & Type\\
\hline
1 & $I_{34}$(3)=14.0 & $r_{34}$(3)=4.135 & $z_{34}$(3)=2, $z_{43}$(3)=2 & 
Fe3 - Fe4\\
2 & $I_{34}$(4)=5.6 & $r_{34}$(4)=4.172 & $z_{34}$(4)=2, $z_{43}$(4)=2 & 
Fe3 - Fe4\\
3 & $I_{14}$(2)=4.2 & $r_{14}$(2)=4.201 & $z_{14}$(2)=3, $z_{41}$(2)=1 & 
Fe1 (dumbbell) - Fe4 (corrugated layer)\\
4 & $I_{23}$(2)=-7.9 & $r_{23}$(2)=4.038 & $z_{23}$(2)=4, $z_{32}$(2)=2 & 
Fe3 - Fe2 (corrugated layer)\\
5 & $I_{33}$(3)=--10.7 & $r_{33}$(3)=3.605 & $z_{33}$(3)=1 & 
Fe3 - Fe3\\
6 & $I_{33}$(5)=--13.2 & $r_{33}$(3)=4.257 & $z_{33}$(5)=2 & 
Fe3 - Fe3\\
7 & $I_{12}$(2)=--14.6 & $r_{12}$(2)=4.108 & $z_{12}$(2)=3, $z_{21}$(2)=2 & 
Fe1 (dumbbell) - Fe2 (corrugated layer)\\
8 & $I_{34}$(5)=--16.3 & $r_{34}$(5)=4.220 & $z_{34}$(5)=2, $z_{43}$(5)=2 & 
Fe3 - Fe4 (corrugated layer)\\
9 & $I_{22}$(2)=--18.1 & $r_{22}$(2)=4.260 & $z_{22}$(2)=4 & 
Fe2 (corrugated layer) - Fe2 (corrugated layer)\\
10& $I_{33}$(4)=--29.0 & $r_{33}$(4)=4.239 & $z_{33}$(4)=2 & 
Fe3 - Fe3 (layer above and below)\\
11& $I_{22}$(1)=--111.1 & $r_{22}$(1)=4.182 & $z_{22}$(1)=2 & 
Fe2 - Fe2 (layer above and below)\\
\end{tabular}
\end{ruledtabular}
\label{tab3}
\end{table*}

In Table~\ref{tab4} the parameters of exchange interaction 
of the ion Fe$a$ ($a$=1, 2, 3, 4) with the ions from the first 
coordination layer, as well as the exchange parameters $I_{33}(2)$ 
with the Fe3 ions located between the first and second layers, 
in the rhombohedral structure of Gd$_2$Fe$_{17}$ are shown. 
Transferring from the hexagonal to the more compact rhombohedral 
structure of Gd$_2$Fe$_{17}$, the distance in a dumbbell Fe1-Fe1 
decreases from $r_{11}^{hex}(1)$=2.400~$\AA$ 
to $r_{11}^{rh}$(1)=2.385~$\AA$. Where from the exchange parameter 
$I_{11}(1)$ increases from $I_{11}^{hex}(1)$=238.8~K to 
$I_{11}^{rh}(1)$=287.5~K. However, the second largest interaction 
$I_{44}(1)$ in the corrugated hexagon plane and the third 
largest interaction $I_{34}(1)$ between the layers 
of Fe3 and Fe4 are smaller in the rhombohedral structure 
of Gd$_2$Fe$_{17}$: $I_{44}^{hex}(1)$=218.5~K vs. $I_{44}^{rh}(1)$=182.2~K. 
While $r_{44}^{hex}(1)$=2.478~$\AA$ is smaller than 
$r_{44}^{rh}(1)$=2.490~$\AA$, and $r_{34}^{hex}(1)$=2.511~$\AA$ 
smaller than $r_{34}^{rh}(1)$=2.549~$\AA$
$I_{34}^{hex}(1)$=136.1~K 
exceeds $I_{34}^{rh}(1)$=125.9~K. 
Also, in the rhombohedral structure one antiferromagnetic exchange
(with negative value)
appears since $I_{33}^{hex}(1)$=80.4~K 
in the hexagonal structure becomes $I_{33}^{rh}(1)$=--36.5~K 
in the rhombohedral one, whereas $r_{33}^{hex}(1)$=2.403~$\AA$
will be enhanced to $r_{33}^{rh}(1)$=2.466~$\AA$.
Similar results are obtained for rhombohedral Y$_2$Fe$_{17}$
in Ref.~\onlinecite{Sabiryanov98}.

\begin{table*}
\caption{Parameters of exchange in the rhombohedral structure 
of Gd$_2$Fe$_{17}$ for the ions of the first coordination sphere 
and indirect exchange $I_{33}(2)$.}
\begin{ruledtabular}
\begin{tabular}{ccccc}
N & Exchange (K) & Distance ($\AA$) & Number of neighbors & Type\\
\hline
1 & $I_{11}$(1)=287.5 & $r_{11}$(1)=2.385 & $z_{11}$(1)=1 & 
Fe1 (dumbbell) - Fe1 (dumbbell)\\
2 & $I_{44}$(1)=182.2 & $r_{44}$(1)=2.490 & $z_{44}$(1)=2 & 
Fe4 (corrugated plane) - Fe4 (corrugated plane)\\
3 & $I_{34}$(1)=125.9 & $r_{34}$(1)=2.551 & $z_{34}$(1)=$z_{43}$(1)=2 & 
Fe3 - Fe4 (corrugated plane)\\
4 & $I_{33}$(2)=121.9 & $r_{33}$(2)=3.563 & $z_{33}$(2)=1 & 
Fe3 - Fe3\\
5 & $I_{24}$(1)=121.0 & $r_{24}$(1)=2.448 & $z_{24}$(1)=4, $z_{42}$(1)=2 & 
Fe2 (corrugated layer) - Fe4 (corrugated layer)\\
6 & $I_{34}$(2)=105.7 & $r_{34}$(2)=2.613 & $z_{34}$(2)=$z_{43}$(2)=2 & 
Fe3 - Fe4 (corrugated layer)\\
7 & $I_{14}$(1)=88.8 & $r_{14}$(1)=2.639 & $z_{14}$(1)=3, $z_{41}$(1)=1 & 
Fe1 (dumbbell) - Fe4 (corrugated layer)\\
8 & $I_{23}$(1)=87.1 & $r_{23}$(1)=2.423 & $z_{23}$(1)=4, $z_{32}$(1)=2 & 
Fe3 - Fe2 (corrugated layer)\\
9 & $I_{12}$(1)=83.6 & $r_{12}$(1)=2.602 & $z_{12}$(1)=3, $z_{21}$(1)=2 & 
Fe1 (dumbbell) - Fe2 (corrugated layer)\\
10& $I_{13}$(1)=74.1 & $r_{13}$(1)=2.740 & $z_{13}$(1)=6, $z_{31}$(1)=2 & 
Fe1 (dumbbell) - Fe3\\
11& $I_{33}$(1)=--36.5 & $r_{33}$(1)=2.466 & $z_{33}$(1)=2 & 
Fe3 - Fe3\\
\end{tabular}
\end{ruledtabular}
\label{tab4}
\end{table*}

\begin{table*}
\caption{Parameters of exchange in the rhombohedral structure 
of Gd$_2$Fe$_{17}$ for the ions of the second coordination sphere.}
\begin{ruledtabular}
\begin{tabular}{ccccc}
N & Exchange (K) & Distance ($\AA$) & Number of neighbors & Type\\
\hline
1 & $I_{14}$(2)=12.3 & $r_{14}$(2)=4.206 & $z_{14}$(2)=3 & 
Fe1 (dumbbell) - Fe4 (corrugated layer)\\
2 & $I_{23}$(2)=1.6 & $r_{23}$(2)=4.063 & $z_{23}$(2)=4,$z_{32}$(2)=2 & 
Fe2 (corrugated layer)- Fe3\\
3 & $I_{34}$(3)=-1.2 & $r_{34}$(3)=4.132 & $z_{34}$(3)=$z_{43}$(2)=2 & 
Fe3 - Fe4 (corrugated layer)\\
4 & $I_{12}$(2)=-3.6 & $r_{12}$(2)=4.095 & $z_{12}$(2)=3,$z_{21}$(2)=2 & 
Fe1 (dumbbell) - Fe2 (corrugated layer)\\
5 & $I_{34}$(5)=-5.0 & $r_{34}$(5)=4.190 & $z_{34}$(5)=$z_{43}$(5)=2 & 
Fe3 - Fe4 (corrugated layer)\\
6 & $I_{13}$(2)=-15.2 & $r_{13}$(2)=4.237 & $z_{13}$(2)=6,$z_{31}$(2)=2 & 
Fe1 (dumbbell) - Fe3\\
7 & $I_{34}$(4)=-16.4 & $r_{34}$(4)=4.171 & $z_{34}$(4)=$z_{43}$(4)=2 & 
Fe3  - Fe4 (corrugated layer)\\
8 & $I_{33}$(3)=-20.4 & $r_{33}$(3)=4.192 & $z_{33}$(3)=2 & 
Fe3 - Fe3 \\
9 & $I_{44}$(2)=-36.7 & $r_{44}$(2)=3.841 & $z_{44}$(2)=1 & 
Fe4(corrugated layer)  - Fe4 (corrugated layer)\\
10& $I_{44}$(3)=-57.5 & $r_{44}$(3)=4.236 & $z_{44}$(4)=2 & 
Fe4(corrugated layer)  - Fe4 (corrugated layer)\\
\end{tabular}
\end{ruledtabular}
\label{tab5}
\end{table*}

Finally, Table~\ref{tab5} contains exchange parameters for the
second coordination sphere of rhombohedral Gd$_2$Fe$_{17}$.
As in the case of hexagonal phase most of them are antiferromagnetic and
considerably weaker in comparison with the first coordination layer.

From these data it is clear that a decrease of the distance 
in a pair of interacting ions in the first coordination sphere results in an increase 
of the ferromagnetic exchange parameter. But since the distance 
in the basic structural elements changes in opposite way from 
the hexagonal to rhombohedral structure (the distance decreases 
in the dumbbells, and mostly increases in the hexagons), then 
it is reasonable to use the Curie point $T_C$ to estimate 
collective effect of these changes in the distances 
and corresponding exchange parameters transferring from one 
structure of Gd$_2$Fe$_{17}$ to another. Let us use Weiss 
mean-field theory taking into account only the Fe-Fe exchange 
interactions.

Within the Weiss mean-field theory the
Hamiltonian~(\ref{Hamiltonian}) can be written as
\begin{equation}
H^{MF}=\frac{1}{2} \sum_{a, b}\sum_{\mathbf{l}_a, 
\mathbf{\Delta}_b} I_{ab}(\mathbf{\Delta}_b)
<S^Z_a(\mathbf{l}_a)> <S^Z_b(\mathbf{l}_a 
+ \mathbf{\Delta}_{ab})>-\sum_{a}\sum_{\mathbf{l}_a} 
h_a(\mathbf{l}_a)S^Z_a(\mathbf{l}_a),
\label{HamiltonianMF}
\end{equation}
where $\mathbf{\Delta}_{ab}=\mathbf{n}_b-\mathbf{l}_a$, 
and molecular field $h_a(\mathbf{l}_a)$ for spin 
$\mathbf{S}_a(\mathbf{l}_a)$ can be presented as 
\begin{equation}
h_a(\mathbf{l}_a)=\sum_{b, \mathbf{\Delta}_{ab}} 
I_{ab}(\mathbf{\Delta}_{ab})<S^Z_b(\mathbf{l}_a
+\mathbf{\Delta}_{ab})>=\sum_{b,|\mathbf{\Delta}_{ab}|} I_{ab}
(|\mathbf{\Delta}_{ab}|)z_{ab}(|\mathbf{\Delta}_{ab}|)\sigma_b.
\label{ha}
\end{equation}
Here $\sigma_b=<S^Z_b(\mathbf{l}_a+\mathbf{\Delta}_{ab})>$ 
denotes the thermodynamic average of $z$-projection 
of classical spin for the ion Fe$b$ ($b$=1, 2, 3, 4).
$z_{ab}(|\mathbf{\Delta}_{ab}|)$ 
is the number of Fe$b$ neighbors at the distance 
$|\mathbf{\Delta}_{ab}|$ from the central ion Fe$a$.

Calculating the average value $\sigma_a$ in a field $h_a$, 
one obtains a set of self-consistent equations 
for $\sigma_a$ ($a$=1, 2, 3, 4):

\begin{equation}
\sigma_a=S_aL(\frac{h_aS_a}{k_BT})
=S_a[coth(\frac{h_aS_a}{k_BT})-\frac{k_BT}{h_aS_a}]
\label{sigmaA}
\end{equation}
($L(x)=coth(x)-1/x$ -- Langevin function). 
Linearizing the set Eq.~(\ref{sigmaA}) with respect 
to small $\sigma_a$, one obtains a set of linear equations 
for $T_C$:
\begin{equation}
k_BT_C\sigma_a=\frac{S_a^2}{3}\sum_{b}
\sum_{|\mathbf{\Delta}_{ab}|} 
I_{ab}(|\mathbf{\Delta}_{ab}|)z_{ab}
(|\mathbf{\Delta}_{ab}|)\sigma_b.
\label{setTC}
\end{equation}

Let us solve the set of equations (\ref{setTC}) for the rhombohedral and hexagonal 
structures of Gd$_2$Fe$_{17}$ restricting ourselves 
by using $I_{ab}(|\mathbf{\Delta}_{ab}|)$ and 
$z_{ab}(|\mathbf{\Delta}_{ab}|)$ as the parameters 
of direct exchange and number of neighbors 
in the first coordination layer (Tables~\ref{tab4} 
and \ref{tab2}).
Upon these assumptions the estimations for the Curie point 
for exchanges only from the first coordination layer
are $T_C^{rh}$=429~K and $T_C^{hex}$=402~K. These estimations 
are below the experimental values $T_C^{rh}$=475~K 
(Ref.~\onlinecite{Shen98}) and $T_C^{hex}$=466~K 
(Ref.~\onlinecite{Knyazev06}).
Such a situation is typical for estimations of $T_C$
for ferromagnetic metals in the framework of the
molecular field approximation (c.f. for instance,
estimations for Fe, Co and Ni\cite{Halilov98}).
At the same time these theoretical values grasp a trend
of the Curie temperature growth in transferring from the
rhombohedral to the hexagonal structures in Gd$_2$Fe$_{17}$
what takes place in experiments Refs.~\onlinecite{Shen98,Knyazev06}.
Here we can address also a question how $T_C$ changes in case second
coordination sphere is taken into account. As shown above in Tables \ref{tab2}-\ref{tab5}
next nearest exchanges are predominantly AFM-like but are weaker. Thus one should expect
a bit of lowering of resulting $T_C$.
Indeed, we found $T_C^{rh}$(1st+2nd)=378~K and $T_C^{hex}$(1st+2nd)=353~K.
Nevertheless a trend of the Curie temperature growth from the rhombohedral to the hexagonal structure
is still observed.

But comparing these values one should keep in mind several 
reasons why the calculated results of $T_C$ cannot {\it precisely} 
fit the experiment. 

Firstly, in accord with spin-fluctuation theory of magnetism approximating 
magnetic excitations in metals with the classical Heisenberg Hamiltonian, 
all exchange interactions in metals are long-range, relatively slowly 
decaying, and sign oscillating (similar to Ruderman-Kittel 
interaction).~\cite{Prange79,Mathon83} This conclusion also follows 
from our results for the hexagonal structure of Gd$_2$Fe$_{17}$, 
see Table~\ref{tab3} (for the rhombohedral structure see Table~\ref{tab5}),
where the exchange interaction of the Fe ions 
with the neighbors of the second-order coordination layer mostly change 
their sign and become negative, antiferromagnetic. Exact account 
of these oscillating interactions requires either exact functional 
form of the exchange distance dependence for a particular pair of ions 
(e.g., Fe1-Fe3), or calculation of the exchange parameters for such 
a pair on the scale of a few oscillations. At the moment the method 
of Ref.~\onlinecite{Liechtenstein84} allows one to calculate the exchange 
parameters, but it is not clear how to present an explicit functional 
dependence of the exchange on distance. Nevertheless, in principle 
in this method one can calculate exchange parameters on the scale 
of a few oscillations, but for a metallic magnet with several types 
of magnetic moments and complicated crystal structure it is 
a cumbersome problem.

Secondly, also according to spin-fluctuation theory in metals,~\cite{Prange79} 
the exchange parameters of the classical Heisenberg Hamiltonian 
in the general case are temperature dependent. For this reason, 
the exchange parameters at $T$=0 and $T$=$T_C$ differ, whereas 
in our calculations we estimated $T_C$ from the exchange parameters 
calculated using the $T$=0 method of Ref. \onlinecite{Liechtenstein84}.

Thirdly, an obvious limitation of such calculations is the Weiss mean-field 
theory, restricted in its accuracy.

Before we summarize let us also compare magnetic properties of
elemental $bcc$ Fe to ones averaged over first coordination sphere
of Fe ions of hexagonal Gd$_2$Fe$_{17}$.
First of all value of nearest neighbor exchange interaction of $bcc$ Fe
(393.2~K,~Ref.~\onlinecite{Lichtenstein01}; 373.6~K,~Ref.~\onlinecite{Pajda01})
is four times larger than the above mentioned averaged value of
Gd$_2$Fe$_{17}$~(99.6~K, Ref.~\onlinecite{avex}).
It can be understood from difference of the nearest neighbor distance
of $bcc$ Fe (2.482~\AA) and averaged distance of
Gd$_2$Fe$_{17}$~(2.574~\AA, Ref.~\onlinecite{avdist}).
Thus hexagonal Gd$_2$Fe$_{17}$ has more loose-packed crystal structure
than $bcc$ Fe. This more loose-packed crystal structure also leads
to a smaller saturation magnetization at $T$=0 of Fe ion sublattice in hexagonal Gd$_2$Fe$_{17}$
(1299~G,~Ref.~\onlinecite{avmag}) in contrast to $bcc$ Fe (1740~G,~Ref.~\onlinecite{Kittel}).
It becomes even more clear since values of $bcc$ Fe local magnetic
moment (2.2~$\mu_B$)
is almost equal to one averaged over all crystallographic positions
for hexagonal Gd$_2$Fe$_{17}$ (2.15~$\mu_B$).
Similar results can be obtained for rhombohedral phase of Gd$_2$Fe$_{17}$.

\section{Conclusion}
\label{conclusion}

In this work we calculate from first principles
electronic structure and values of exchange 
interaction parameters of the Fe ions sublattice
for the hexagonal and rhombohedral phases
of intermetallic compound Gd$_2$Fe$_{17}$
within first and second coordination spheres.
Based on the values of exchange interaction paremeters
Curie temperateres $T_C$ were also calculated.
Obtained theoretical Curie temperatures $T_C^{rh}$=429~K and $T_C^{hex}$=402~K are found to be slightly
below the experimental values $T_C^{rh}$=475~K~\cite{Shen98} and $T_C^{hex}$=466~K~\cite{Knyazev06}.
However the tendency of $T_C^{rh}$ to be higher than $T_C^{hex}$ is well reproduced.
The same is found for the case if second coordination sphere is taken into account.
Corresponding Curie temperatures are $T_C^{rh}$(1st+2nd)=378~K and $T_C^{hex}$(1st+2nd)=353~K.
 
Beside that from our calculations we observe
that for the first coordination sphere exchange between
different types of Fe ions is ferromagnetic (with only
one exception for rhombohedral phase).
At the same time exchange 
for next nearest neighbors is mostly antiferromagnetic one.
Latter one is consistent with RKKY exchange picture
(exchange sign depends on pair bond length).
\cite{Hubbard79,Prange79,Wang82,Mathon83}
However such competition only is not enough
to decrease significantly Curie temperature $T_C$ of $R_2$Fe$_{17}$
as proposed if Ref.~\onlinecite{Givord74}.
Also it was reported before that exchange value between Fe
in dumbbell positions is the determinating factor for the $T_C$ value\cite{Sabiryanov98}.
However in this paper we showed that there are several exchange interactions
almost of the same strength as dumbbell one, e.g. $I_{44}$ (see Tab.~\ref{tab2}).

Let us also roughly analyze values of exchange interactions in terms of crystal structure.
Collecting data from Tables~\ref{tab2}-\ref{tab5} one can observe for both phases analogous tendency.
Exchanges within any layer are weaker than interlayer ones.
Exceptional is only Fe2 ion surrounded with four Fe4 ions
with exchange $I_{24}(1)$. Anisotropy of direct exchange interaction
parameters in the first coordination sphere comes from
Fe-3d orbitals overlap anisotropy. On top of that
RKKY exchange gives slight modulations in the first coordination
sphere because of complexity of the Fermi surface of the compounds.
For the second coordination sphere RKKY is main contribution
to exchange interaction but for the reason mentioned above
is also highly anisotropic.

After all we propose another one observation to explain
relatively low $T_C$ of R$_2$Fe$_{17}$ series.
To this end we compare magnetic properties of Fe ion sublattice of
hexagonal Gd$_2$Fe$_{17}$ to elemental $bcc$ Fe.
There are two ingredients influencing $T_C$ value: local magnetic moment values
and exchange interaction values.
Despite the fact that for both hexagonal Gd$_2$Fe$_{17}$ and $bcc$ Fe
local magnetic moments of Fe ion are almost identical,
exchange interaction values of $bcc$ Fe is four times stronger
then for hexagonal Gd$_2$Fe$_{17}$\cite{avex}.
It comes from more dense packing of Fe ions of $bcc$ Fe
compared to hexagonal Gd$_2$Fe$_{17}$\cite{avdist}.
The same arguments support lower saturation magnetization value for Gd$_2$Fe$_{17}$.
Thus even for highest $T_C$ compound Gd$_2$Fe$_{17}$ of $R_2$Fe$_{17}$ series
magnetic properties are found to be inferior to ones of $bcc$ Fe.

\section{Acknowledgments} 
We thank M.A. Korotin and V.I. Anisimov for helpful discussions.
This work was supported by the Russian Foundation for Basic Research, 
(project nos. 08-02-00021, 08-02-00712, and 09-02-00431), the Presidium 
of the Russian Academy of Sciences (RAS) ``Quantum macrophysics'' 
and of the Division of Physical Sciences of the RAS ``Strongly correlated 
electrons in semiconductors, metals, superconductors and magnetic materials''. 
A.L. acknowledges ``Dynasty'' Foundation. 
I.N. acknowledges Russian Science Support Foundation and MK-2242.2007.2.

\newpage

\newpage

\begin{thebibliography}{99}
\bibitem{Gd_radii}Indeed, for the hexagonal phase of Gd$_2$Fe$_{17}$ there are two
inequivalent types of Gd atoms. From calculating of Gd atomic sphere radii through merging of Hartree potentials
one can obtain R(Gd1)=2.86~a.u. and R(Gd2)=3.89~a.u. which are quite different from each other.
Since Gd atoms do not play leading role in our computation we chose for simplicity
R(Gd1)=R(Gd2)=2.86~a.u. by introducing additional empty spheres. It does not affect significantly
in its turn Fe atomic sphere radii which are more important here.
\bibitem{2nd_layer}Since distribution of atoms becomes more uniform with distance
there is quite difficult to determine unambiguously upper border of second coordination layer.
Thus for both phases we chose it to be 4.3~\AA. This value corresponds to the beginning of
the second thin out spherical layer.
\bibitem{avex}Averaged over first coordination sphere
exchange interaction value for hexagonal Gd$_2$Fe$_{17}$ is computed as
$\bar I=\frac{I_{11}(1)z_{11}(1)+I_{12}(4)z_{12}(1)+...+I_{44}(4)z_{44}(1)}
{z_{11}(1)+z_{12}(1)+...+z_{44}(1)}=99.6$~K,
where $I_{ab}$ and $z_{ab}$ are taken from Tab.~\ref{tab2}.
\bibitem{avdist}Averaged over first coordination sphere
Fe-Fe distance for hexagonal Gd$_2$Fe$_{17}$ is computed as
$\bar r^{hex}=\frac{r_{11}(1)z_{11}(1)+r_{12}(4)z_{12}(1)+...+r_{44}(4)z_{44}(1)}
{z_{11}(1)+z_{12}(1)+...+z_{44}(1)}=2.574$~\AA,
where $r_{ab}$ and $z_{ab}$ are taken from Tab.~\ref{tab2}.
\bibitem{avmag}To get saturation magnetization at $T$=0 of Fe ion
sublattice in hexagonal Gd$_2$Fe$_{17}$ we took magnetic moment
per formula unit M($\Sigma$Fe)=36.54~$\mu_B$ and unitary cell volume
$\Omega$=521.4$\times$10$^{-24}$~cm$^3$ (Ref.~\onlinecite{Knyazev06}).
\end{thebibliography}
\end {document}